\documentclass[10pt, twocolumn, nofootinbib,superscriptaddress]{revtex4-1}
\usepackage{amsmath,amssymb,amsfonts}
\usepackage{algorithmic}
\usepackage{graphicx}
\usepackage{textcomp}
\usepackage{xcolor}
\usepackage{ragged2e}
\usepackage{booktabs, makecell, tabularx}
\usepackage{lipsum}

\usepackage{gensymb}
\usepackage{lineno}
\makeatletter
    \renewcommand\@make@capt@title[2]{%
     \@ifx@empty\float@link{\@firstofone}{\expandafter\href\expandafter{\float@link}}%
      {\textbf{#1}}\@caption@fignum@sep#2\quad}%
\makeatother
\makeatletter 
\renewcommand{\fnum@figure}{\textbf{Fig.~\thefigure}} 
\makeatother
\usepackage{xcolor}
\newcommand{\beginsupplement}{%
        \setcounter{table}{0}
        \renewcommand{\thetable}{S\arabic{table}}%
        \setcounter{figure}{0}
        \renewcommand{\thefigure}{S\arabic{figure}}%
     }
\def\BibTeX{{\rm B\kern-.05em{\sc i\kern-.025em b}\kern-.08em
    T\kern-.1667em\lower.7ex\hbox{E}\kern-.125emX}}

\begin{document}

\author{Riley te Morsche}
\affiliation{Nonlinear Nanophotonics Group, MESA+ Institute of Nanotechnology,\\
University of Twente, Enschede, Netherlands}
\author{Lukas Puts}
\author{Weiming Yao}
\affiliation{Photonic Integration Group, Department of Electrical Engineering,\\
Eindhoven University of Technology, Eindhoven, Netherlands}
\author{David Marpaung}
\email{david.marpaung@utwente.nl}
\affiliation{Nonlinear Nanophotonics Group, MESA+ Institute of Nanotechnology,\\
University of Twente, Enschede, Netherlands}

\date{\today}
\title{Giant backward Brillouin interaction in generic InP integrated photonics}

\begin{abstract}
We report the first measurement of backward stimulated Brillouin scattering (SBS) in generic InP waveguides, supporting enhanced Brillouin gain of $g_B/Q_m = 3.5 \pm 0.6~\text{W}^{-1} \text{m}^{-1}$ mediated by weakly-guided pressure waves in InGaAsP. This leads to SBS gain coefficients as high as $737 \pm 54~\text{W}^{-1} \text{m}^{-1}$, observed in a mature, foundry-accessible photonic integration platform.

\end{abstract}
\maketitle
\section{Introduction}

Stimulated Brillouin scattering (SBS) is a third-order nonlinear interaction between optical photons and GHz acoustic waves. It is well-known for limiting the maximum power handling of long fibre lines, more recently being explored in photonic integrated circuits. In backward Brillouin scattering, an incident (pump) photon is elastically scattered from a co-propagating phonon, and Doppler shifted to a lower (higher) frequency, creating a counter-propagating Stokes (anti-Stokes) photon. The Stokes process involves the emission of phonons and supports self-sustained oscillation, hence the naming stimulated Brillouin scattering \cite{BoydNonlinearOptics}. This permits efficient optical energy transfer over a bandwidth defined by the phonon lifetime, which is typically on the order of 10-100s of MHz.

The narrowband and directional nature of SBS has attracted much research attention, for instance in integrated microwave photonic signal processing \cite{Marpaung2015} and generation \cite{Merklein2016}, optical isolation \cite{Poulton2012, Otterstrom2019a}, spectrum analysis \cite{Wang2026}, distributed sensing \cite{Zarifi2019} and optical gyroscopes \cite{Lai2020}. In addition, the three-wave dynamics in SBS exhibits phase noise filtering when the lifetimes of the optical and acoustic fields are very different \cite{Debut2000}, enabling ultra-low fundamental linewidth on-chip lasers \cite{Gundavarapu2019, Liu2025}. 

Many of these achievements have been obtained using established integrated Brillouin photonics platforms, including chalcogenides (ChG) \cite{Morrison2017, Song2021}, silicon-on-insulator (SOI) \cite{Lei2024} and silicon nitride ($\text{Si}_3\text{N}_4$) \cite{Gundavarapu2019, Botter2022, Klaver2024}. However, these face challenges such as material toxicity and photodarkening \cite{Zhang2021}, complex fabrication and waveguide suspension \cite{Lei2024}, or low gain coefficients \cite{Gundavarapu2019}. Recently, thin-film lithium niobate (TFLN) attracts much attention for its versatile nonlinearities \cite{Boes2023} and features moderate SBS gain \cite{Rodrigues2023, Ye2025b}. Tantalum pentoxide ($\text{Ta}_2\text{O}_5$) has been proposed for CMOS-compatible, low-loss nonlinear photonics \cite{Brodnik2026}, with SBS gain an order of magnitude larger than $\text{Si}_3\text{N}_4$ \cite{Liu2026}. Finally, SBS has also been observed in compound-semiconductor-on-insulator (CSOI) platforms, namely AlGaAs-OI \cite{Jin2026} and InGaP-OI \cite{Xue2026cleo}. While the portfolio of Brillouin-active waveguides has greatly expanded in recent years, transitioning from lab to real-world environments remains challenging, usually limited by propagation loss requirements, CMOS-compatibility or ability to monolithically integrate other system components like efficient modulators and photodetectors, light sources and nonlinear frequency conversion. 

Indium phosphide (InP) is a semiconductor material widely used in the photonics industry to fabricate, for example, commercial optical transceivers. Here, we investigate SBS in InP waveguides, fabricated on a foundry-accessible platform \cite{Smit2014, Augustin2018}. This mature technology platform has already been used to fabricate complex photonic integrated circuits, comprising many passive components \cite{Kleijn2014thesis}, polarisation control \cite{Dzibrou2014thesis, Mehrabi2026}, tunable lasers \cite{Latkowski2015} and high-speed modulators \cite{Hillier2025}. This flexibility has led to scalable demonstrations of emerging technologies like THz generation \cite{Li2025} and photonic neural network components \cite{Puts2025}.

In this paper, we expand the InP toolbox by adding backward SBS. Optical waveguides in InP are formed in $\text{In}_{1-x}\text{Ga}_x\text{As}_{1-y}\text{P}_y/\text{InP}$, with $x = 0.47-0.47y$ the equation for lattice matching to the InP substrate. Tuning the composition $y$ and n/p-type doping changes optical, electronic and acoustic properties. We show that near-constant stiffness but increasing density with GaAs content \cite{Adachi2009} results in higher-order pressure wave confinement similar to optical fibre \cite{Kobyakov2010, Poulton2013} and large mode area SOI \cite{Ye2025a}. This leads to large SBS gain coefficients, up to $g_B = 737~\text{W}^{-1}\text{m}^{-1}$ at a frequency shift $\Delta\nu_B \approx 20.8~\text{GHz}$. This places InP on-par with ChG in terms of gain coefficient and outperforms other scalable platforms, see Tab.~\ref{tab:sbs_gain_list} for a comparison. The narrow linewidth $\Gamma_B \approx 100$~MHz provides new opportunities for high-performance microwave and optical signal processing in a foundry platform.

\begin{figure*}[t!]
\centering
\includegraphics[width=\linewidth]{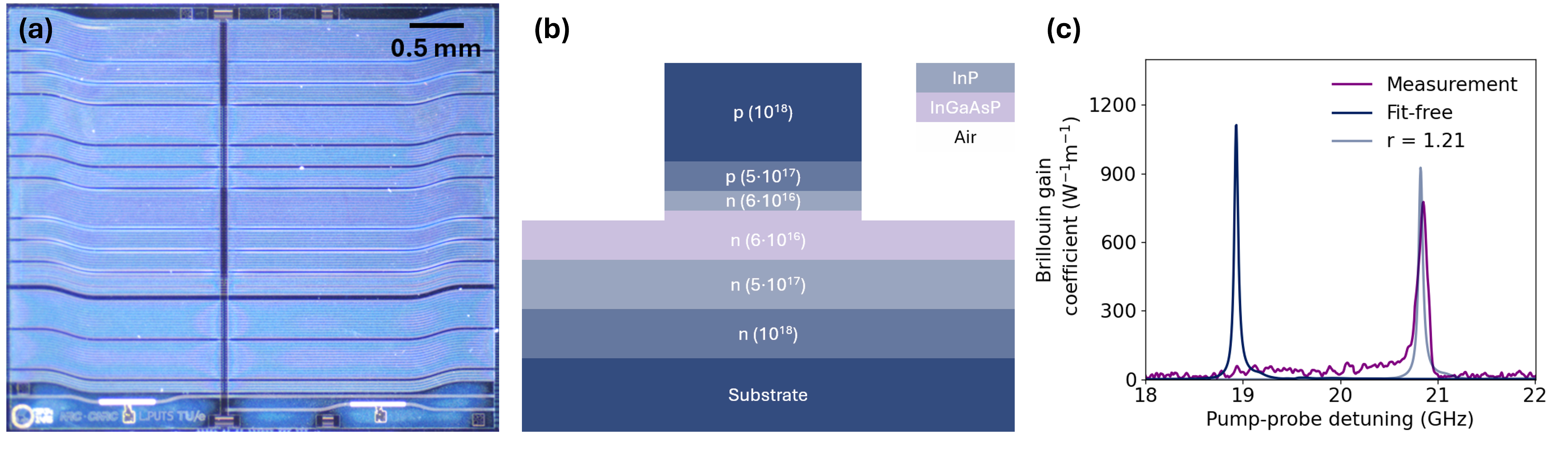}
\caption{\textbf{Stimulated Brillouin scattering in a generic InP waveguide}. \textbf{(a)} Standard $4\times4.5~\text{mm}^2$ InP chip containing passive and active waveguides. \textbf{(b)} shallow-etched waveguide geometry (adapted from \cite{DAgostino2015thesis}). \textbf{(c)} Example calibrated SBS measurement, showing strong SBS gain of around $737~\text{W}^{-1}\text{m}^{-1}$ at 20.85 GHz. Overlaid are gain spectra obtained from fit-free and stiffness-adjusted (factor 1.21) simulations.}
\label{fig1}
\end{figure*}

\section{Results}

We investigate backwards, intramodal Brillouin scattering in generic InP waveguides produced in a commercial foundry. The devices reported in this work were fabricated by Smart Photonics through the JePPIX multi-project wafer (MPW) service. The chip used in our experiments is shown in Fig.~\ref{fig1}~(a) and consists of many passive and active devices. Two chip samples were used, both featuring two identical, 2-$\mu$m-wide shallow-etch passive waveguides, schematically depicted in Fig.~\ref{fig1}~(b). The InGaAsP waveguide material has a bandgap energy of 1.25~eV and is also denoted Q(1.25~eV). From the bandgap, we know the core composition $y=0.53$ \cite{Dzibrou2014thesis} and are able to calculate other optical and acoustic properties.

We measured the SBS gain spectrum of all waveguides by pump-probe measurement using the intensity-modulation and lock-in detection scheme \cite{Botter2022}. The waveguide gain coefficient is calculated from the fibre response in the same measurement, explained in detail in Supplementary Note~A. Fig.~\ref{fig1}~(c) shows a representative measured Brillouin gain spectrum, obtained from background-removed and normalised lock-in data. The average Brillouin gain coefficient of the corresponding waveguide is $g_B \approx 737\pm54~\text{W}^{-1}\text{m}^{-1}$, frequency shift $\Delta\nu_B \approx 20.85~\text{GHz}$ and linewidth $\Gamma_B \approx 104~\text{MHz}$.

We repeat the measurement to determine experimental uncertainty and sample variations. These statistics are shown in Fig.~\ref{fig3}~(a-c). Here, we plot the calculated waveguide Brillouin gain coefficient $g_B$, linewidth $\Gamma_B$, and normalised gain $g_B/Q_m$. The measured SBS shifts in the two chips show MHz-level uncertainty, being 20.85~GHz (20.78~GHz) for waveguides 1-2 (3-4). From the measured statistics, we conclude that the normalised Brillouin gain coefficient of these InP waveguides is about $g_B = 3.5 \pm 0.6 ~\text{W}^{-1}\text{m}^{-1}$, comparable to $\text{As}_2\text{S}_3$ \cite{Morrison2017}.

\begin{figure}[b!]
\centering
\includegraphics[width=0.9\linewidth]{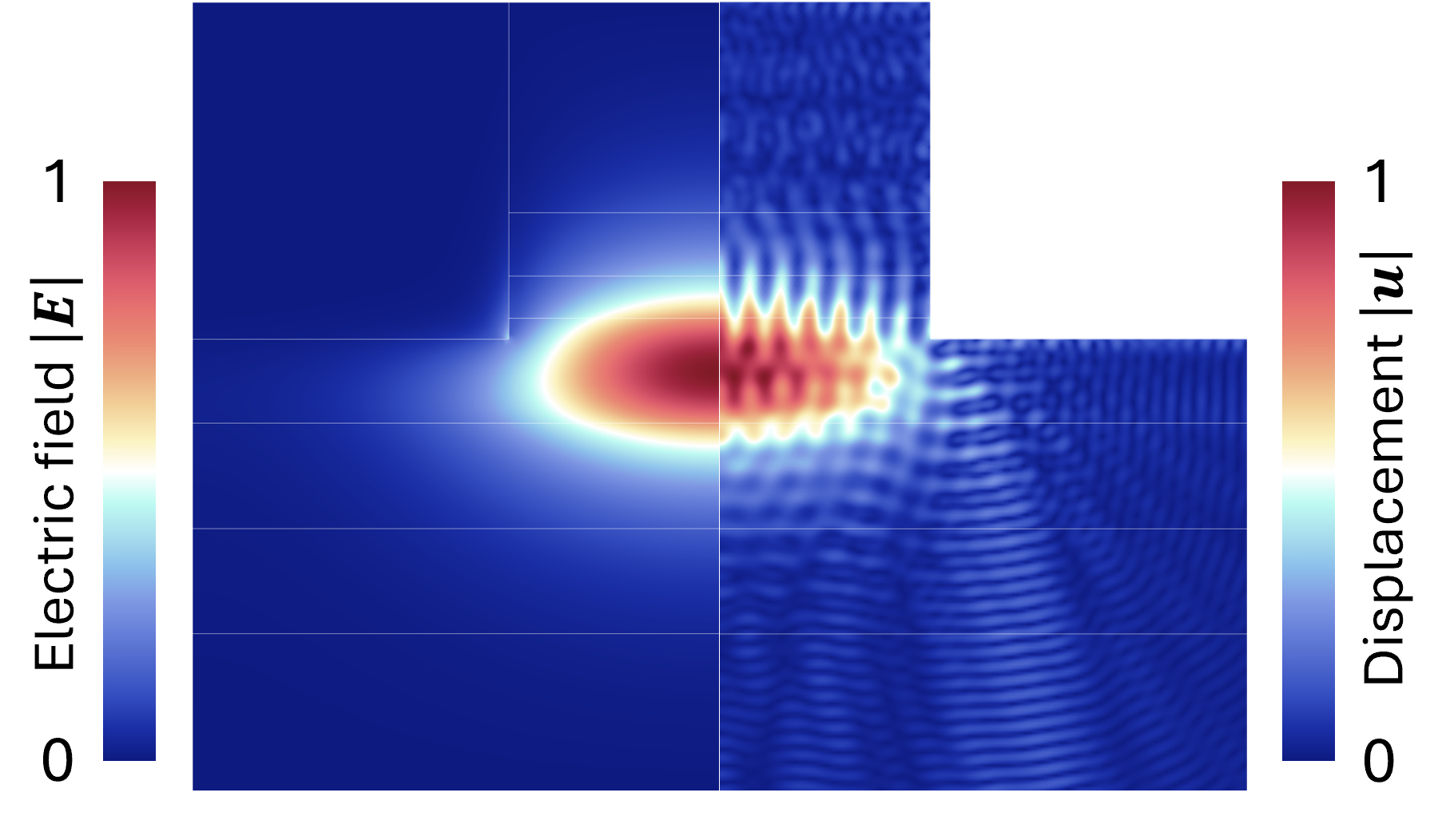}
\caption{\textbf{Brillouin gain simulation}. Left: normalised electric field of the fundamental TE mode with effective index $n_{\text{eff}} = 3.2518$, right: normalised steady-state displacement response of the fit-free gain peak at 18.935 GHz in Fig.~\ref{fig1}~(c).}
\label{fig2}
\end{figure}

\begin{figure*}[t!]
\centering
\includegraphics[width=\linewidth]{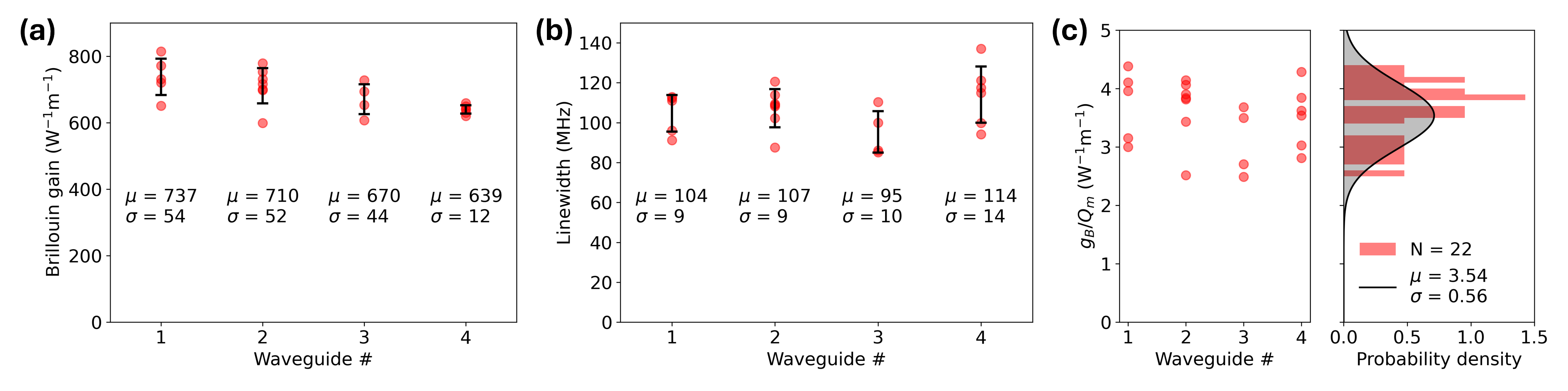}
\caption{\textbf{Brillouin gain statistics from all measurements.} \textbf{(a)} Calibrated peak SBS gain coefficient for all waveguide samples, showing all data, their mean $\mu$ and standard deviation $\sigma$. \textbf{(b)} Brillouin gain linewidths determined from Lorentzian fitting. \textbf{(c)} normalised gain coefficients $g_B/Q_m$ for all measurements (left) and combined histogram (right).}
\label{fig3}
\end{figure*}

To understand the physical origin of the strong SBS response of the InP waveguide, we conduct numerical simulations of the waveguide Brillouin gain spectrum. In Fig.~\ref{fig2} we show the simulated TE-like optical mode and acoustic response at the Brillouin frequency shift, being around 18.935~GHz in the fit-free simulation. In principle, the shallow-etch rib waveguide structure is not ideally suited to total internal reflection confinement due to slab radiation. However, the high refractive index $n_{\text{eff}} \approx 3.25$ leads to a very small phonon wavelength $q \approx 240~\text{nm}$ for backward SBS. The relatively large optical mode compared to acoustic wavelength leads to weak guidance of higher-order pressure waves. The steady-state displacement response at the Brillouin gain centre is also plotted in Fig.~\ref{fig2} and shows strong similarity with the (transverse) electric field. This produces a large optomechanical overlap through $p_{12}$ ($\approx -0.242$ \cite{Otterstrom2023}). Combined with the high index ($n_{\text{eff}}\approx3.25$) this produces impressive Brillouin gain of $g_B^{\text{ES}} = 1110~\text{W}^{-1}\text{m}^{-1}$, dominated by electrostriction\footnote{The radiation pressure contribution is entirely negligible in our simulation.}. The simulated gain linewidth is slightly smaller than the measurement at around 49~MHz, corresponding to the mechanical quality factor $Q_m \approx 386$. The larger measured linewidth may be explained by different factors, like underestimated absorption loss or neglecting acoustic scattering due to sidewall/interface roughness. There may also be some inhomogeneous broadening, e.g. from waveguide inhomogeneity. The normalised gain coefficient in simulation is $g_B^{\text{ES}}/Q_m = 2.88~\text{W}^{-1}\text{m}^{-1}$, somewhat lower than the measurement.


There is a significant discrepancy between the measured and fit-free simulated SBS shifts, being around 20.8~GHz and 18.9~GHz, respectively. To better fit the experiment, we conducted a simulation with a uniform $\times1.21$ increased stiffness, to emulate an increase in effective acoustic mode stiffness. In this simulation, the SBS shift increases due to the higher acoustic velocity, shifting up to 20.825~GHz closely matching the experiment. The peak Brillouin gain is lower at $920~\text{W}^{-1}\text{m}^{-1}$ as expected from the inverse relationship between SBS gain and acoustic mode stiffness \cite{Wiederhecker2019, Rodrigues2025}. As the mechanical quality is unaffected, the normalised gain reduces to $g_B^{\text{ES}}/Q_m = 2.38~\text{W}^{-1}\text{m}^{-1}$.

While our current simulations provide an understanding of the basic mechanism of the observed SBS process, the frequency difference and lower gain coefficient in fit-free simulations indicate that there might be a difference between assumed material properties and those of the physical sample. From our estimation, the difference in frequency shift cannot be explained by changes in material properties from n/p-doping. Aside from the acousto-optic properties, some physical process may be overlooked, like piezoelectricity, though we are currently unable to confirm the exact mechanism. It has previously been noted that piezomechanics need to be included to accurately predict Brillouin frequency shifts in TFLN \cite{Rodrigues2025}, generally causing an increase in acoustic mode stiffness. InP also features a piezo response, though notably weaker than LN ($r_{14}=-1.80~\text{pm/V}$ for InP \cite{Adachi2005} compared to $r_{33} = 33~\text{pm/V}$ for LN \cite{Sinatkas2021}). Alternatively, there may be carrier-induced modulation of the Brillouin response \cite{Otterstrom2023}, potentially caused by free-carrier absorption, also warranting further study. 

In summary, we observed strong backward SBS interaction in standard InP waveguides, expanding the nonlinear toolbox of mature and foundry-accessible generic InP. This opens up many research directions, potentially enabling low-noise Brillouin lasers, likely operating in the phonon lasing regime \cite{Otterstrom2019b}. Well-established applications like microwave photonics could benefit from the narrow SBS gain linewidth, circumventing strict propagation loss requirements. This works provides a starting point for monolithic Brillouin photonics systems in InP, benefitting from existing foundry building blocks, and potential for electronic interfacing via small form-factor pluggable or co-packaged optics. 

\section{Methods}

All experimental details are provided in Supplementary Note~A. In fitting fibre and waveguide Brillouin signal voltages, we take into account the phase between background and SBS signals and the lock-in amplifier reference. Not taking this into account in these measurements would lead to overestimation of the waveguide Brillouin gain coefficient. A complex fitting procedure is used to reliably extract the gain coefficient, and we use measurement statistics to estimate the uncertainty in the Brillouin gain resonances and the best-estimate $g_B/Q_m$ of the shallow-etched InP waveguide. 

Simulations are conducted using the formalism outlined in \cite{Rakich2012}. Simulation details and a list of material properties are provided in Supplementary Note~B. Where available, we use literature values for the material properties of $\text{In}_{1-x}\text{Ga}_x\text{As}_{1-y}\text{P}_y/\text{InP}$. The refractive index of the core $n \approx 3.36$ is estimated using the known composition $y=0.53$, consistent with a theoretical model in \cite{Adachi2009} and the expected effective index around 3.25 \cite{Smit2014}. Density and stiffness are linearly interpolated over composition using reported endpoint values. Due to limited available experimental data, we opted to use photo-elastic coefficients calculated for $y=0.625$ by Otterstrom et al. \cite{Otterstrom2023}. 

\bigskip
\section*{Author Contribution}

R.M. and D.M. developed the concept and proposed experiments. R.M. designed and conducted experiments with input from D.M., and performed data analysis. R.M. developed simulations with input from L.P. and W.Y. L.P. designed and provided InP samples. R.M. wrote the manuscript with input from all authors. D.M. supervised the project. 

\label{sec:four}

\begin{acknowledgments}

The authors acknowledge funding from the European Research Council Consolidator Grant (101043229 TRIFFIC) and the Photon Delta National Growth Fund programme.

Nazca Design was used to generate the mask layout in this work.

\end{acknowledgments}

\section*{Disclosures}
The authors declare no conflicts of interest.

\section*{Data Availability}
The data of this study are available from the corresponding authors upon reasonable request.

\bibliographystyle{IEEEtran}
\bibliography{library}

\newpage
\onecolumngrid
\beginsupplement
\newpage
\section{Supplementary Note A: Experimental details}

\textit{I. Loss characterisation}

\noindent
Before measuring the SBS response of the InP waveguides, we measure the insertion loss and nonlinear (2PA, FCA) loss. We test two chips (14H5, 14H9) each featuring two passive shallow-etched waveguides, see Fig. 1 (a). Linear insertion loss, measured fibre-to-fibre through the chip for this set of samples varies between $8 – 10$~dB. The estimated propagation loss for this waveguide is 3~dB/cm, so per facet coupling loss is around 4~dB. The insertion loss increases with pump power, starting around 20~dBm off-chip, adding up to 7~dB excess loss at 27~dBm off-chip, which is the highest power that was used. We take 20~dBm as the maximal pump power, minimising nonlinear loss to 1~dB without significant loss in signal strength. The probe laser is kept at low power (3~dBm) to reduce probe-induced background oscillations, and amplified using a low-noise EDFA before photodetection. Thermal and shot noise in these measurements were found negligible compared to background signal and data fitting uncertainties.\\

\textit{II. Intensity-modulation lock-in measurement}

\begin{figure*}[b!]
\centering
\includegraphics[width=\linewidth]{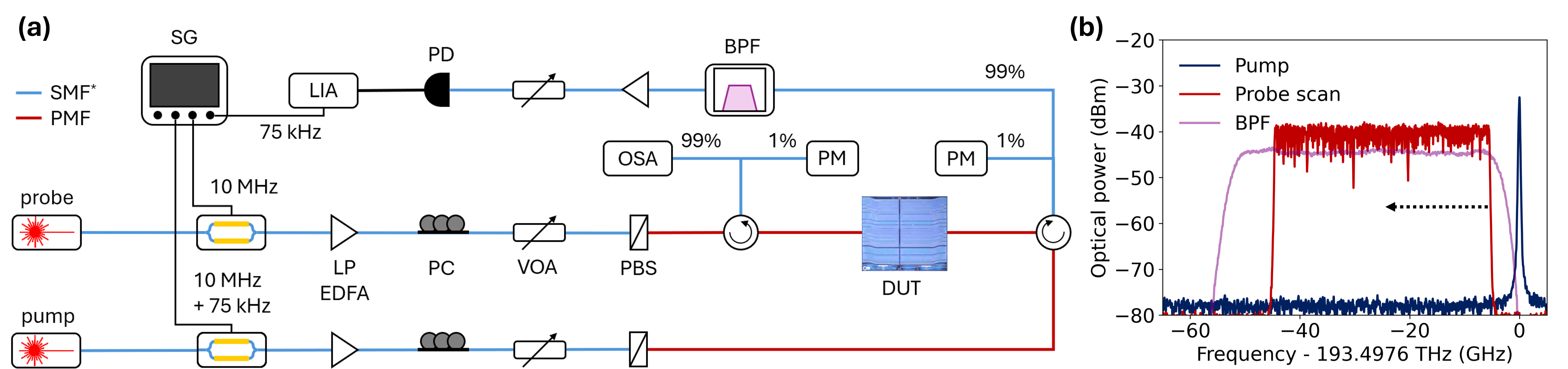}
\caption{\textbf{Intensity-modulated pump-probe lock-in amplifier setup.} \textbf{(a)} Setup schematic, where intensity-modulated pump and probe lasers are injected in counter-propagating directions through the device under test (DUT). As the probe laser is scanned down in frequency, the pump-induced SBS gain, oscillating at the baseband frequency of 75~kHz, is recorded on the lock-in amplifier (LIA). \textbf{(b)} Optical spectrum analyser (OSA) recordings of the pump laser, max-hold of the probe being scanned down in frequency, and transmission profile of the bandpass filter (BPF). Abbreviations: SMF/PMF: single-mode/polarisation-maintaining fibre, SG: signal generator, PC: polarisation controller, VOA: variable optical attenuator, PM: power meter, PD: photodiode.}
\label{fig_s1new}
\end{figure*}

\noindent
We measured SBS gain in the InP waveguides using lock-in detection with the intensity modulated pump-probe setup \cite{Botter2022}. The setup is shown in Fig.~\ref{fig_s1new}~(a). In this experiment, a pump and probe laser are intensity modulated at 10.075~MHz and 10~MHz, respectively, and injected into the waveguide sample in counter-propagating directions. The probe is tuned downwards in frequency, and experiences amplification due to the strong pump light in the fibres and waveguide, at characteristic frequencies corresponding to Brillouin gain resonances. Reflected pump light is filtered out using a bandpass filter, after which the probe is amplified using a low-noise EDFA and fed to the photodetector. The lock-in amplifier records the pump-induced Brillouin amplification of the probe, which has a component oscillating at the baseband frequency of 75~kHz due to the differential modulation. We continuously monitor the pump and probe lasers using a high-resolution optical spectrum analyser, producing a frequency axis calibrated to around 20~MHz accuracy. The ratio of fibre and waveguide SBS signal amplitudes allows calculation of the waveguide Brillouin gain coefficient \cite{Gyger2020, Botter2022}:
\begin{equation} \label{eq_s1}
    g_{B,\text{wg}} = \frac{V_{\text{wg}}}{V_{\text{PMF}}} \frac{g_{B,\text{PMF}} P_{\text{pump}} L_{\text{PMF}} }{ P_{\text{pump,wg}} P_{\text{eff,wg}}}
\end{equation}

Because of the relatively low insertion loss of $9 - 11$~dB, there is some residual pump light in the probe-side PM lensed fibre, resulting in about 10\% additional effective fibre length: $L_{\text{eff,PMF}} = (\text{3~m}) \cdot (1 + 10^{-\text{IL}/10})$. The on-chip pump power is calculated using the per-facet coupling loss, estimated as $\text{FL} = \frac{1}{2} (\text{IL} - \alpha L_{\text{wg}})$. The waveguide and fibre voltages $V_{\text{wg}}$ and $V_{\text{PMF}}$ are extracted from the data using a complex fitting procedure, including background signals explained in the following section. In this procedure we allow rotations of the background/SBS signals in the complex plane, since the lock-in amplifier also records phase. If the phase is ignored, the fibre peak appears too small because of the background, leading to overestimated waveguide gain coefficients. \\

\newpage
\textit{III. Origin of background signals}

\begin{figure*}[t!]
\centering
\includegraphics[width=0.9\linewidth]{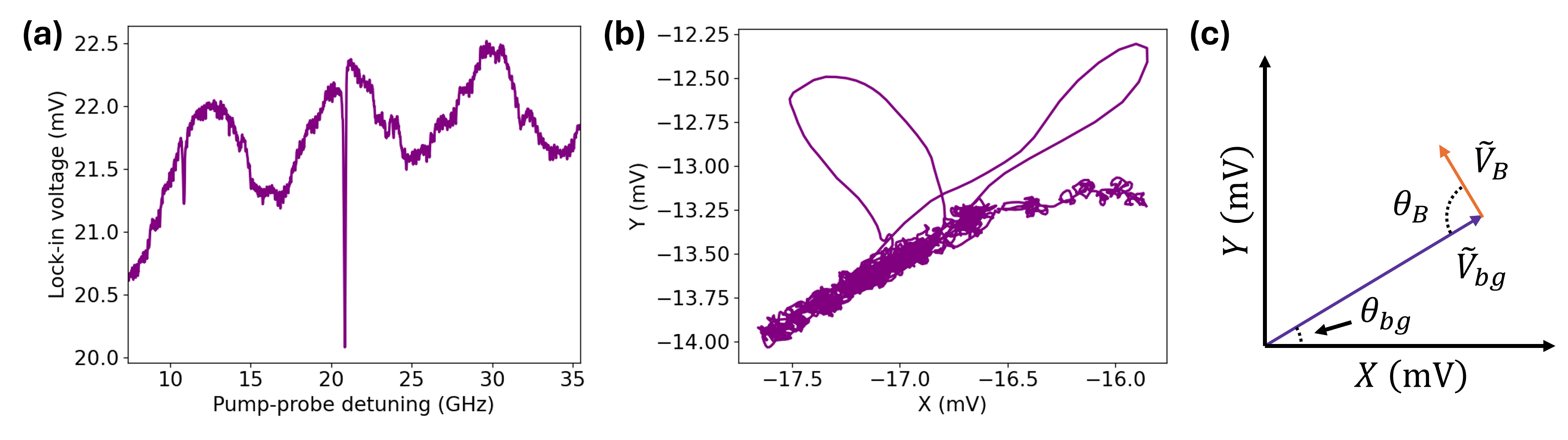}
\caption{\textbf{Intensity-modulation lock-in amplifier measurement.} \textbf{(a)} Example of the full measured spectrum, showing fibre (waveguide) Brillouin peaks at around 10.8 (20.8) GHz pump-probe detuning, on a strong background signal, with a sinusoidal component varying with the waveguide FSR of about 8.5 GHz. \textbf{(b)} Phasor plot of the LIA voltage, showing the oscillating background and distinct lobes of the fibre (waveguide) Brillouin signals at 90\degree (180\degree) offset from the background. \textbf{(c)} Sketch of the lock-in amplifier voltage as the sum of background ($\tilde{V}_{bg}$) and Brillouin signal ($\tilde{V}_B$) phasors.}
\label{fig_s1}
\end{figure*}

\noindent
It is known that Raman- and Kerr-nonlinearities can interfere with the measurement of Brillouin gain in this setup \cite{Gyger2020}. The strong pump light modulates the waveguide length, which in turn affects its transmission response through the Fabry-Pérot response through the reflections at the waveguide facets. This pump-induced Kerr-FP modulation imparts the pump modulation frequency onto the probe, generating a background signal in the LIA baseband. In our setup, an additional, smaller background component is generated by FP-induced sideband imbalance on the probe laser while it is scanned down in frequency. In Fig.~\ref{fig_s1}~(a) we show the lock-in amplifier voltage over the full measured spectral range. Fibre and waveguide Brillouin signals are visible at 10.8~GHz and 20.8~GHz. The narrow SBS signals are easily distinguished from the strong backgrounds, which show a large constant component due to pump-induced FP modulation, and a smaller sinusoidal component due to probe sideband imbalance varying with the waveguide SFR of around 8.5~GHz. Fig.~\ref{fig_s1}~(b) shows the same data plotted in the complex plane, where the two SBS peaks show up as distinct lobes offset with a fixed angle relative to the background. Fig.~\ref{fig_s1}~(c) shows a phasor diagram explaining this signal as the sum of two components, indicating angles for the background voltage $\theta_{\text{bg}}$ and difference of background and SBS voltage $\theta_B$. These angles are fixed, since the lock-in voltage is generated by beating with external local oscillator, phase-locked to the pump and probe modulator signals.

\begin{figure*}[b!]
\centering
\includegraphics[width=0.9\linewidth]{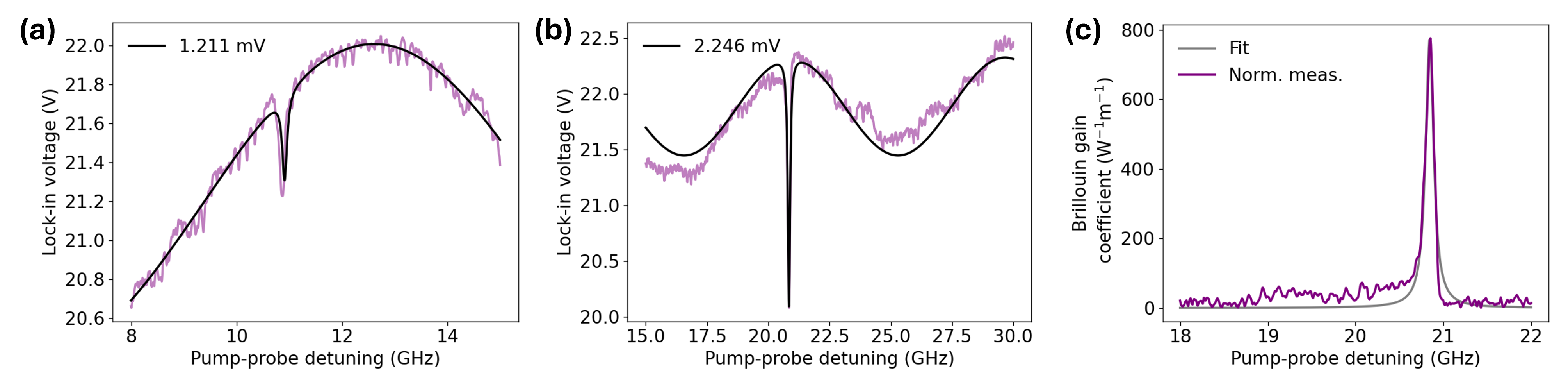}
\caption{\textbf{Calibration of the waveguide Brillouin gain coefficient.} \textbf{(a)} Zoom-in on the full spectrum of Fig. II.1 showing the fibre peak near 10.8 GHz and complex Lorentian fit. The legend shows signal amplitude in the phasor plane. \textbf{(b)} Same as (a), for the waveguide Brillouin signal near 20.8 GHz. \textbf{(c)} Background-removed and normalised Brillouin gain measurement, for the example shown in Fig.~\ref{fig1}~(c) in the main text.}
\label{fig_s2}
\end{figure*}

It is established in the literature that the Kerr-FP background signal can be remedied using an auxiliary pump laser, also known as the triple-intensity modulation technique \cite{Gyger2020, Ye2024}. An auxiliary pump laser is introduced, detuned far from the SBS pump and modulated 180\degree out-of-phase to equalise the power coupled into the waveguide in the time domain. If done correctly, this cancels out Kerr-induced FP oscillations and Raman scattering-induced background from both the waveguide and fibre in the setup. However, when applying this technique here, we found that the pump-induced FP modulation in these waveguides is so strong that the background could not be adequately reduced without compromising signal strength. Fortunately, without using the auxiliary pump, the distinct Brillouin gain features and stable phases of  background and Brillouin signals still allow us to extract fibre and waveguide SBS voltages, provided we include the phase in fitting. Doing so, we are able to reliably extract the waveguide Brillouin gain coefficient in the usual manner \cite{Botter2022}. We neglect the RF modulation factor \cite{Ye2024}, which is negligible for the Brillouin linewidths measured here. \\

\textit{IV. Gain calibration and uncertainty analysis}

\noindent
The fit function used to process lock-in data is:
\begin{equation}
    V_{\text{LIA}}(f) = V_{\text{bg}} \left( 1 + a \cos(f/\text{FSR} + \varphi_{\text{0}}) \right) e^{i \phi_{\text{bg}}} + V_{\text{B}} \mathcal{L}(f, \Delta\nu_B,\Gamma_B) e^{i \phi_{\text{B}}}
\end{equation}

Here, $V_{\text{bg}}$, $V_{\text{B}}$ are background and Brillouin voltage amplitudes, $a$ accounts for the sinusoidal background varying with waveguide FSR and $\varphi_0$ phase offset, $\phi_{\text{bg}}$, $\phi_{\text{B}}$ are the phases of background and Brillouin signals, and $\mathcal{L}(f, \Delta\nu_B,\Gamma_B) = (\Gamma_B/2) \cdot ( (f - \Delta\nu_B)^2 + (\Gamma_B/2)^2 )^{-1}$ is the Lorentzian line shape of the Brillouin gain. Fig.~\ref{fig_s2}~(a-b) show example fits of the fibre and waveguide signals. The black curves show the fit, including a constant plus sinusoidal background and Lorentian component for the Brillouin voltage. Both background and SBS signal are allowed to rotate in the complex plane. The legend indicates the real magnitude of the SBS signals, different than the apparent magnitude due to phase rotations. In particular, the fibre SBS voltage $V_{B,\text{PMF}} \approx 1.211~\text{mV}$ is notably larger than its apparent magnitude in Fig.~\ref{fig_s2}~(a). Using these fit values and Eq.~\ref{eq_s1}, the Brillouin gain spectrum can be calculated by removing the fit background and normalising the gain peak to the calculated $g_{B,\text{wg}}$, shown in Fig.~\ref{fig_s2}~(c) for the measurement in the main text (Fig.~\ref{fig1}~(c)). The fit shows the normalised waveguide Brillouin gain spectrum $g_{B} \mathcal{L}(f, \Delta\nu_B,\Gamma_B)$, which overlaps well with the waveguide response.

We estimate the error in $g_B/Q_m$ due to experimental uncertainties by repeating the lock-in measurements. For each sample, we collected $N = 5 - 8$ measurements and calculate the gain coefficient. Fig.~\ref{fig3}~(a) in the main text shows the calibrated peak gain $g_B$ for all measurements and waveguide samples. These values are based on Eq.~\ref{eq_s1} with parameters listed in Tab.~\ref{tab:exp_data}. We take the standard deviation of all values as the error. Waveguides 1-2 (3-4) are on the same chip, labeled 14H5 (14H9). We observe a slight decrease in peak gain in the second chip, potentially due to wafer-level variations in fabricated waveguide geometry, or optical/acoustic properties. Fig.~\ref{fig3}~(b) shows the linewidths obtained from the Lorentzian fit, showing consistent values close to 100~MHz. We found neglible variations in SBS shift in the data, being 20.85~GHz (20.78~GHz) for chip 14H5 (14H9). Using the shifts and linewidths, we calculate the mechanical quality factor $Q_m = \Gamma_B /\Delta\nu_B$. The mean waveguide gain coefficient across all samples is $g_B/Q_m = 3.5 \pm 0.6~\text{W}^{-1}\text{m}^{-1}$, see Fig.~\ref{fig3}~(c).

\begin{table}[h]
    \centering
    \setlength\tabcolsep{6pt} 
    \begin{tabular}{llll}
        \textbf{Parameter} & \textbf{Value} & \textbf{Unit} & \textbf{Description} \\
         \hline
        $P_{\text{pump}}$ & 20 &~dBm & Off-chip pump power \\
        $P_{\text{probe}}$ & 3 &~dBm & Off-chip probe power \\
        $P_{\text{PD}}$ & 8 &~dBm & Amplified probe power arriving on photodiode \\
        $g_{B,\text{PMF}}$ & 0.18 & $\text{W}^{-1}\text{m}^{-1}$ & PMF Brillouin gain coefficient \\
        $L_{\text{PMF}}$ & 3 & m & PMF length between chip and circulator \\
        $L_{\text{wg}}$ & 4.5 & mm & InP waveguide length \\
        $\text{IL}$ & $9 - 11$ &~dB & Range of measured insertion loss \\
        $\alpha$ & 3 & $\text{dB}~\text{cm}^{-1}$ & Estimated waveguide propagation loss \\
         \hline
    \end{tabular}
    \caption{\textbf{Experimental parameters used to calibrate the waveguide gain coefficient.}}
    \label{tab:exp_data}
\end{table}

\section{Supplementary Note B: Material properties and simulation details}

\textit{I. Brillouin gain model}

\noindent
We simulate the Brillouin gain spectrum using COMSOL Multiphysics, using electrostriction and radiation pressure forces to the simulate steady-state displacement response, and calculating the Brillouin gain from their overlap integral according to \cite{Rakich2012}. The geometry is taken from \cite{DAgostino2015thesis}, though we neglect the influence of doping and incorporate only two materials, being undoped InP and Q(1.25~eV) InGaAsP. Both materials have a cubic/wurtzite crystal structure and the InP substrate has the [100] crystal orientation. All waveguides are directed predominantly along cartesian axes, so we do not rotate the crystal w.r.t. the waveguide propagation direction. Note that acoustic confinement (and corresponding the Brillouin gain spectrum) along non-cartesian directions will be different because P-wave velocities of both materials depend on propagation direction w.r.t. the crystal axes \cite{Adachi2009}. This likely results in some background gain response, but this is not resolved in our measurements. The waveguide bends present in our sample could also contribute to the lower simulated $g_B/Q_m$ value compared to the measurement estimate.

In the fit-free simulation, we adopt optical and acoustic parameters known from the literature on InP and InGaAs/InP \cite{Adachi2009, Carlotti1996, Dzibrou2014thesis} and linearly interpolate values to Q(1.25~eV). Photoelastic constants of InGaAsP show a complex dependence on composition, and we opted to use values from previous work on acousto-electric Brillouin scattering in InGaAsP \cite{Otterstrom2023}. While calculated for $y=0.625$, the slight difference in compositions with our work does not meaningfully impact the outcome of the simulations. The simulated SBS gain is additionally dominated by $p_{12}$, so the shear gain coefficient can be simply rescaled given updated knowledge of the $p_{ij}$ of Q(1.25~eV). Mechanical absorption is fixed to $Q_{\text{abs}} = 1000$. The fit-free simulation shows a strong Brillouin gain response centered at 18.935~GHz, with a peak gain coefficient of around $1110~\text{W}^{-1}\text{m}^{-1}$ and linewidth of 49~MHz. The corresponding quality factor is $Q_m = 386$, leading to a normalised gain $g_B^{\text{ES}} = 2.88~\text{W}^{-1}\text{m}^{-1}$, not too dissimilar but lower than the measured value. However, the frequency shift is notably different from the experiment, being roughly 10\% lower.

In the simulation, we first calculate the fundamental optical TE-like mode, shown in Fig.~\ref{fig2}~(a) in the main text. The effective index sets the acoustic wavevector used to calculate the displacement response. The displacement response is calculated in the steady-state for varying acoustic frequency, being generated by the phase-matched photoelastic stress due to the beating of forwards (pump) and backward (Stokes) propagating optical fields. The peak displacement response, at a frequency shift of 18.935~GHz, is show in Fig.~\ref{fig2}~(b). We find strong Brillouin gain mediated by a higher-order, weakly guided pressure wave. The large mode profile compared with the acoustic wavelength $q \approx 240~\text{nm}$ increases the mechanical quality factor, behaviour also seen in thick SOI waveguides \cite{Ye2025a}. The displacement profile also shows strong overlap with the (transverse) electric field, resulting in a high optomechanical overlap through the $p_{12}$ photoelastic component, which is large compared to other scalable SBS platforms. This combined with reasonably small mode area and $n^7$ refractive index scaling of SBS gain \cite{Kobyakov2010}, explains the strong Brillouin gain observed in these waveguides.\\

\textit{II. Material properties}

\noindent
While many optical and acoustic properties of InP and InGaAsP semiconductor alloys are known, not all parameters required for Brillouin gain modelling are known perfectly. In addition, the effects of doping type and concentration are not well known. In the simulation, we make minimal assumptions to be able model SBS in the $\text{In}_x \text{Ga}_{1-x} \text{As}_{1-y} \text{P}_y/\text{InP}$ quaternary. The lattice matching equation is $x = 0.47 y - 0.47$ \cite{Adachi2009}, and the core composition is $y = 0.53$, corresponding to a bandgap of 1.25~µm \cite{Dzibrou2014thesis}. The refractive index of the core material is estimated from the composition according to a theoretical model in Ch.~10 of \cite{Adachi2009}, and leads to an expected effective index of about 3.25 \cite{Smit2014}. Errors in the estimated refractive index are expected to be below 1\%, and hence cannot account for the difference in the simulated and measured Brillouin shifts, being 18.93~GHz and 20.85~GHz, respectively. 

We linearly interpolate the density and stiffness according to values reported in the literature \cite{Adachi2009, Carlotti1996}. Because of a complicated dependence on material composition, as well as limited experimental data being available, we adopt the same photoelastic coefficients from previous theoretical work on Brillouin scattering in InGaAsP \cite{Otterstrom2023}. Fortunately, the SBS interaction in this work is completely dominated by the $p_{12}$ coupling. The complete set of properties is listed in Tab.~\ref{tab:mat_data}.

We opt to neglect doping concentration on material properties, which is justified at least for the refractive index, since the expected variation is below 0.4\% for $p = 10^{18}~\text{cm}^{-3}$, and much lower for the other layers where most of the optical field is concentrated \cite{Dzibrou2014thesis}. Assuming a fixed lattice constant, the effects of dopants on the density is expected on the order of $10^{-4}$. To the best of our knowledge, there is no published data on the change in stiffness with doping type and/or concentration.

\begin{table}[h]
    \centering
    \setlength\tabcolsep{6pt} 
    \begin{tabular}{lllllll}
        \textbf{Material} & \textbf{InP} & \textbf{InGaAsP} & \textbf{InGaAs} & \textbf{Unit} & \textbf{Description} & \textbf{Ref.}\\
        \textbf{property} & ($y=1$) & ($y=0.53$) & ($y=0$) & & & \\
         \hline
        $n$ & 3.169 & 3.36* & - & 1 & Refractive index & \cite{Dzibrou2014thesis} \\
        $\rho$ & 4790 & 5142 & 5531 & $\text{kg}~\text{m}^{-3}$ & Density & \cite{Adachi2009, Carlotti1996} \\
        $C_{11}$ & 102.2 & 100.2 & 98.0 & GPa & Stiffness component & \cite{Adachi2009, Carlotti1996} \\
        $C_{12}$ & 57.3 & 54.7 & 51.8 & GPa &  &  \\
        $C_{44}$ & 44.2 & 45.5 & 46.9 & GPa &  &  \\
        $p_{11}$ & -0.127 & $-0.127^{\dagger}$ & - & 1 & Photoelastic component & \cite{Otterstrom2023} \\
        $p_{12}$ & -0.109 & $-0.242^{\dagger}$ & - & 1 &  &  \\
        $p_{44}$ & -0.064 & $0.027^{\dagger}$ & - & 1 &  &  \\
         \hline
    \end{tabular}
    \caption{\textbf{Acousto-optic material properties used in simulations.} Mechanical properties of the InGaAsP core are linearly interpolated to the composition $y=0.53$, based on the designed bandgap of 1.25 µm \cite{Dzibrou2014thesis} *Estimated from expected effective index around 3.25 \cite{Smit2014}.$~^{\dagger}$Interpolated photo-elastic constants for $y=0.625$ adopted from previous work \cite{Otterstrom2023}.}
    \label{tab:mat_data}
\end{table}

\section{Supplementary Note C: Comparison with existing platforms}

Tab.~\ref{tab:sbs_gain_list} shows a comprehensive list of backward SBS gain, achieved in scalable photonic integration platforms, being either using CMOS-compatible materials or requiring only one back-end-of-line production step. In this list, a few platforms stand out for their high gain coefficient, being suspended SOI and AlGaAs/InGaP-on-insulator. In silicon, large gain is achieved by constructive interference of electrostrictive and radiation pressure forces. Small mode areas and high refractive index contrast further boosts the gain, but suspension is required to confine acoustic waves, due to silicon's large stiffness compared to silica. On the other hand, AlGaAs/InGaP both feature inherent acoustic confinement on silica, and produce large gain dominated by electrostriction. In the intermediate gain category are TFLN and $\text{Te}\text{O}_2$-coated $\text{Si}_3\text{N}_4$ waveguides. TFLN features inherently large photoelastic coefficients and acoustic guidance particularly for Rayleigh waves. The gain is lower for x-cut waveguides, which is often used because of its benefits for electro-optic modulators. $\text{Si}_3\text{N}_4$-$\text{Te}\text{O}_2$ has been proposed as a single back-end-of-line step to introduce strong Brillouin gain into the otherwise passive waveguide platform. This approach makes use of the strong photoelastic effect and low acoustic velocity in $\text{Te}\text{O}_2$, which unlike other chalcogenide glasses does not suffer from photodarkening or volatility. Dilute and double-stripe $\text{Si}_3\text{N}_4$ waveguide platforms suffer from low gain, but compensate by the potential for ultra-low propagation loss, bringing them closer to fibre in scope of Brillouin applications. However, achieving this low loss is a significant challenge for fabrication and usually requires high-temperature annealing, exceeding CMOS limits. Recently, $\text{Ta}_2\text{O}_5$ was shown to have an order of magnitude higher gain than $\text{Si}_3\text{N}_4$, while also featuring ultra-low loss, low thermo-optic and stronger Kerr nonlinearity.

Among all these platforms, InP stands out for being the most versatile featuring lasers, amplifiers, modulators and photodetectors, and being CMOS-compatible and accessible in volume from foundries with mature production facilities. In this work we add backward Brillouin scattering to this extensive library of functionalities, with gain coefficient among the highest currently reported in the literature.



\begin{table}[h]
    \centering
    \begin{tabular*}{\linewidth}{@{\extracolsep{\fill}} llllllllll} 
        \textbf{Platform} & \textbf{Optical} & \textbf{Ac. mode/} & $\mathbf{g_B}$ & $\mathbf{\Delta\nu_B}$ & $\mathbf{\Gamma_B}$ & $\mathbf{Q_m}$ & $\mathbf{g_B/Q_m}$ & $\mathbf{\alpha}$ & \textbf{Ref.} \\
         & \textbf{modes} & \textbf{polarisation} & ($\text{W}^{-1}\text{m}^{-1}$) & (GHz) & (MHz) &  & ($\text{W}^{-1}\text{m}^{-1}$) & ($\text{dB}~\text{m}^{-1}$) & \\
         \hline
        Suspended SOI & TE-TE & P & 600 & 19 & 14 & 1300 & 0.46 & 30 & \cite{Lei2024} \\
        Thick SOI & TE-TE & P & 2.5 & 37.6 & 81.5 & 461 & 0.0054 & 6.7 & \cite{Ye2025a} \\
         \hline
        TFLN z-cut & TE-TE & SAW & 84.9 & 8.14 & 70 & 116 & 0.73 & 80 & \cite{Ye2025b} \\
         & TE-TM & SAW & 80 & 7.38 & 17 & 437 & 0.18 & 112-419 & \cite{Rodrigues2025} \\
        TFLN x-cut & TE-TE & SAW & 29.3 & 8.6 & 20 & 430 & 0.068 & 20 & \cite{Ye2025b} \\
         \hline 
        Dilute $\text{Si}_3\text{N}_4$ & TE-TE & P & 0.1 & 10.9 & 153 & 71 & 0.0014 & 0.4 & \cite{Gundavarapu2019} \\
        SDS $\text{Si}_3\text{N}_4$ & TE-TE & P & 0.38 & 12.9 & 130 & 99 & 0.0038 & 19 & \cite{Botter2022} \\
        $\text{Si}_3\text{N}_4$-$\text{Te}\text{O}_2$ & TE-TE & SAW & 81 & 8.1 & 17 & 476 & 0.17 & 74 & \cite{Klaver2024} \\
         \hline
        $\text{Ta}_2\text{O}_5$ & TE-TE & P & 4.9 & 11.23 & 100 & 112 & 0.044 & 80 & \cite{Liu2026} \\
         \hline
        AlGaAs-OI & TE-TE & SH & 30 & 10.2 & 12.5 & 816 & 0.037 & 28 & \cite{Jin2026} \\
         & & SV & 150 & 12.5 & 17.5 & 714 & 0.21 &  & \\
         & & P & 100 & 20.8 & 150 & 139 & 0.72 &  & \\
        InGaP-OI & TE-TE & SAW & 588 & 9.347 & 5.2 & 1800 & 0.33 & 244 & \cite{Xue2026cleo} \\
         \hline
        \textbf{InGaAsP/InP} & \textbf{TE-TE} & \textbf{P} & \textbf{737} & \textbf{20.85} & \textbf{104} & \textbf{200} & \textbf{3.5} & \textbf{300} & \textbf{This work} \\
         \hline
    \end{tabular*}
    \caption{\textbf{Comparison of backward SBS in scalable photonic integration platforms.} All data are at telecom pump wavelengths of around 1550~nm. Acoustic modes/polarisations are categorised as: P (pressure), SAW (surface acoustic wave), SH/SV (shear horizontal/vertical).}
    \label{tab:sbs_gain_list}
\end{table}

\end{document}